\documentclass[fleqn,usenatbib]{mnras}

\usepackage{newtxtext,newtxmath}

\usepackage[T1]{fontenc}
\usepackage{multicol}
\DeclareRobustCommand{\VAN}[3]{#2}
\let\VANthebibliography\thebibliography
\def\thebibliography{\DeclareRobustCommand{\VAN}[3]{##3}\VANthebibliography}

\usepackage{graphicx}	
\usepackage{amsmath}	

\title[He Abundance in NGC\,1850]{The He abundance in NGC\,1850 A and B: are we observing the early stage of  formation of multiple populations in a stellar cluster?\thanks{This paper is dedicate to the memory of Dr. Antonio Sollima, who passed away prematurely in 2023.}}

\author[R. Carini et al.]{
R. Carini,$^{1}$\thanks{E-mail: roberta.carini@inaf.it}
A. Sollima $^2$
E. Brocato,$^{1,3}$
K. Biazzo $^{1}$
\\
$^{1}$INAF - Osservatorio Astronomioco di Roma (OAR), via Frascati 33, 00078, Monte Porzio Catone (RM), Italy \\
$^{2}$ INAF-Osservatorio di Astrofisica e Scienza dello Spazio (OAS), Via Giobetti 93/3, 40129 Bologna, Italy \\
$^{3}$INAF - Osservatorio Astronomico d'Abruzzo, via M. Maggini snc, I-64100 Teramo, Italy 
}

\date{Accepted for publication by MNRAS}

\pubyear{2023}

\begin{document}
\label{firstpage}
\pagerange{\pageref{firstpage}--\pageref{lastpage}}
\maketitle


\begin{abstract}
We present the result of a sample of  B-stars in the Large Magellanic Cloud young double stellar cluster NGC\,1850 A and NGC\,1850 B, observed with the integral-field spectrograph at the Very Large Telescope, the Multi Unit Spectroscopic Explorer.
We compare the observed equivalent widths (EWs) of four He lines (4922 \AA, 5015 \AA, 6678 \AA, and 7065 \AA) with the ones determined from synthetic spectra computed with different He mass  fraction (Y=0.25, 0.27, 0.30 and 0.35) with the code SYNSPEC, that takes into account the  non-LTE effect.
From this comparison, we determined the He mass fraction of the B stars, finding a  not homogeneous distribution. 
The stars can be divided in three groups, He-weak (Y< 0.24) and  the He-normal (0.24$\leqslant$Y$\leqslant$ 0.26) belonging to the MS of NGC 1850 A, and the He-rich stars (0.33$\leqslant$Y$\leqslant$0.38) situated in the MS associated to NGC 1850 B. 
We have analyzed  the stellar rotation as possible responsible of the anomalous features of the He lines in the He-rich stars. We provide a  simple analysis  of the differences between the  observed EWs and the ones obtained from the theoretical models with different rotation velocity (V$sin$ $i$ = 0 and 250 Km/s). The resolution of the MUSE spectra do not allow to get a conclusive result, however our analysis  support the  He-enhanced hypothesis.

\end{abstract}

\begin{keywords}
methods:observational -- technique: spectroscopic -- stars: abundances --galaxies:star clusters: individual: NGC 1850
\end{keywords}



\section{Introduction}

NGC\,1850 is a young ($t_{age}$ $\sim$ 90 Myr; \citealt{niederhofer15}) and massive ($M$ $\sim$ 5.5 x $10^4$ $M_{\odot}$; \citealt{fischer93}) stellar cluster in  the Large Magellanic Cloud (LMC), located at the edge of the galaxy bar.
The cluster appears
to be  a binary system (NGC\,1850 A and NGC\,1850 B), and there is an indication of tidal interaction between the system members \citep{fischer93}. NGC\,1850 B is located $\sim$ 30" W of the main cluster, constitutes $\sim$ 2\% of the total cluster population, end it is characterized by a significant younger age (between 4 Myr and 15 Myr, \citealt{fischer93,gilmozzi94,vallenari94,antony22}).
The color-magnitude diagram (CMD) of NGC\,1850 A discloses 
a main sequence turn-off (MSTO) region that is wider than what
is expected from a single stellar population and shows the presence
of two main sequences (MSs): a blue and poorly populated MS, and a red MS
which contains the majority of the stars. These features are  not compatible with the typical photometric uncertainties, like field star contamination or differential reddening.  \citep{bastian16,correnti17,milone18}.

The split of the MS and the extended-MSTO (eMSTO), have been observed  in many young (age less 800 Myr and $\sim$2.5 Gyr respectively) massive star clusters (YMCs) in the Magellanic Clouds (MCs) \citep{milone23}.
Unlike Galactic Globular Clusters (GGCs) in which the multiple populations observational evidences are  mainly caused by star-to-star chemical abundance variations (e.g He, Na, O, see review by \citealt{bastian18} and \citealt{gratton19}, \citealt{milone22} and reference therein), the  YMCs of the MCs appears chemically homogeneous (e.g. \citealt{mucciarelli14,milone20}). The origin of the eMSTOs and the   split MS is still debated. In the early works the eMSTO has been interpreted as the result of a prolonged star formation (e.g \citealt{mac08,keller11}). Nowadays, stellar rotation has been suggested as the dominant cause of these photometric features \citep{bastian09,dantona2015}, however   rotation effects do not fully reproduce the observed MSTOs of the entire sample of the CMD observed  in the MCs (e.g \citealt{milone17}).
As a consequence, some authors suggest that a mix of age variation and rotation could explain in a more effective way the eMSTOs and  split MS  phenomena \citep{goud17,costa19}.
In the case of NGC\,1850, \cite{correnti17} 
suggested that the combination of single stellar populations (SSPs) with an age range of $\sim$ 35 Myr along with different rotation rates could explain explain the eMSTO and the MS split of the cluster.

Recently, \cite{kamann23} found 
rotation velocity differences along the blue and red MSs, 
with the blue arm  primarily  consisting of slow rotators ($\sim$ 100 Km/s),  while the red arm  mainly consists of rapid rotators ($\sim$ 200 km/s).

Until now, the phenomena of multiple stellar population in YMCs and in GGCs seems 
not sharing the same origin and cause.


By adopting the hypothesis of a dynamical association between NGC\,1850 A and NGC\,1850 B,  this binary cluster seems to have the right mass and size to evolve in the next 10 Gyr toward the typical configuration of
a present-day GC formed by two distinct stellar populations \citep{baum18}.  
Thus, it exist the intriguing possibility that this cluster could serve as a unique bridge between the young massive stellar clusters observed during the formation of the multiple populations and the GCs exhibiting multiple populations  observed some Gyr after their  formation.

With the aim of better understanding the stellar populations in YMCs and their possible evolution toward a GC appearance, we  analyze the  chemical abundance in NGC\,1850 A and NGC\,1850 B.

The presence of He spread  has been the main marker of multiple populations in GCs (eg. \citealt{cassisi17,gratton19,milone22} and reference therein).
All the formation scenarios aiming to explain the origin of the multiple populations in GCs, predict  that the second generation of stars are enhanced in helium.
The difference in He abundance between stellar population can explain 
the peculiarities in the CMDs of the GGCs, as the main-sequence split and the morphology of the horizontal branch (see review \citealt{bastian18,gratton19,milone22} and reference therein). 
 In fact, stars with higher helium, due the higher molecular weight, evolve more rapidly, and are more luminous than lower helium models of the same
mass, so the sequence of the He-rich models is bluer.
Moreover, the different scenarios  predict quite different 
He abundances at the end of the formation processes.
For example, in the AGB (Asymptotic Giant Branch) scenario the maximum helium enhanced expected is of the order of 0.36-0.38 in terms of mass fraction  \citep{siess10,doherty14}, differently
the fast-rotating massive star scenario predicts helium values up to $\sim$ 0.8  \citep{chant16}.
Therefore, the He abundance is clearly crucial to constrain the origin of the multiple populations.\\

The first direct spectroscopic measurement of highly He-enhanced stars ( Y $\sim$ 0.34) has been provided by \cite{marino14} analyzing the blue horizontal branch of the  GGC NGC\,2808.


Although helium is the second most abundant element in stars, estimates of helium abundance in stellar cluster are quite rare.
Recently, \cite{lagioia19} found traces of He enhancement ($\delta Y$ $\sim$ 0.01) in the second population of star of four GC belonging to the SMC and showing ages in the range of $\sim$ 6$-$10 Gyr.
\cite{carini20} estimated the He abundance of 10 star in the stellar cluster NGC\,330 in the SMC, finding a mean value of   $\langle$$\epsilon$(He)$\rangle$ = 10.93 $\pm$ 0.05, without  evidence of star-to-star helium abundance difference.
\cite{li23} found an He spread of  $\delta Y$ $\sim$ 0.06-0.07 in the MS dwarf stars in the old GC NGC\,2110, in the LMC.
We intend to tackle this puzzling scenario by
evaluating the helium abundance in NGC\,1850 A and B.  
In particular, we analyze  the He abundance in the B-type stars,  whose visible spectra are dominated by lines of He I and the Balmer series of hydrogen. These are rather hot, massive stars with effective temperatures in the range from 1.0 $\times$ $10^4$ to 3.0 $\times$ $10^4$ K and masses in the range from 2 to 20 $M_{\odot}$.

Here, we continue the line of research of \cite{antony22}, using the same set of data observed with the Multi Unit Spectroscopic Explorer
(MUSE) integral field spectrograph in adaptive optics mode
and the photometric parameters found by the authors  during their analysis.

This paper is organized as follows.
In Sec. \ref{sec1}  we describe the MUSE observations and data reduction, while  the synthetic spectra are illustrated in Sec.  \ref{sec:syns}. The data analysis, the determination of the He abundance, the uncertainties associated  and the influence of the stellar rotation on the He features are presented in Sec. \ref{sec2}. The final remark are provided in the last Section.

\section{Observations}
\label{sec1}
We analyzed a set of cubes observed with the integral field spectrograph MUSE \citep{bacon,kelz16} at the Very Large Telescope(VLT) under the observing program 0102.D-0268(A) (PI: Bastian).  The observations were conducted over a period of 6 nights between January and March 2019. Two fields were observed during each visit, namely a central field located at the cluster center (exposure time $t_{exp}= 2\times400s$)  and an outer field  ($t_{exp}= 3\times500s$) approximately 1 arcmin southeast of the cluster center, with a small overlap with
the central field. The total exposure time per visit was 2300s.
The data were obtained using the wide-field mode, with 
 the ground-layer adaptive optics system activated. Each field of view is of 1 $\times$ 1 arcmin$^2$ with a spatial sampling of 0.2 arcsec. The wavelength range covered by the instrument is $\lambda$ = 4800-9300 \AA\ with a  low-to-medium resolution (R $\sim$ 1700-3500).
 Due to the emission of the laser guide star, the portion of spectrum between 5805 \AA\ and 5965 \AA\ has been masked.
 We downloaded the cubes reduced from the ESO Archive Science Portal\footnote{http://archive.eso.org/scienceportal/home}.
 The six  data cubes,  containing flux and wavelength-calibrated spectrum in each spaxel,  cover NGC\,1850 A and B both visible in the right edge  of the central point. 
 Moreover, it was available the  combined cube.
 This  data set have already been used to determine the binary frequency on each of the arms of the split MS  \citep{kamann21} as well as to study the peculiar system NGC\,1850-BH1 \citep{saracino22, elbadry22}, and the effect of the shell stars on the shape of the MSTO \citep{kamann23}.
\cite{antony22} determined the global metallicity on the basis of individual red supergiant (RSG) spectra ($\left< [M/H] \right> = -0.31 \pm 0.01$), the Ba abundances ( $\left< [Ba/Fe] \right > =0.4 \pm 0.02$) and the dynamical mass ($\log(M/M\odot)$ = 4.84 $\pm$ 0.10) of the cluster.
Moreover, the authors analyzed   the O abundance among bright MS stars in NGC\,1850. They found two groups of stars, the O-strong stars ([O=Fe] = -0.16 $\pm$ 0.05) agrees with the value measured in stars with a similar metallicity
in the LMC bar \citep{swa13}, and the O-weak stars displaying no O I triplet absorption lines. The authors interpreted this bimodality as evidence for different stellar rotation rates.

\section{Synthetic spectra}
\label{sec:syns} 

In order to analyze the B stars spectra, we have computed detailed synthetic spectra 
with
SYNSPEC, version 54 \citep{syns54} using non-LTE (NLTE) line-blancked,   model atmospheres of the grid $BSTAR2006$ \cite{Lanz07} for early  B-stars.
We chose the non-LTE approach because many authors, (e.g \citealt{auer73, pryz05,pryz06}) have shown  in their  analysis  how the LTE approach leads to erroneous abundance of He in B stars.
The model atmospheres, calculated with TLUSTY \citep{Lanz07}, consider the 1D plane–parallel geometry, with hydrostatic and radiative equilibrium.
The grid contains  16 effective temperatures $15000$ K $\leq$ $T_{eff}$ $\leq$ $30000$ K, with $1000$ K steps, 10 surface gravities, $2.50$ $\leq$ $\log g$ $\leq$ $4.75$, with 0.25 dex steps and for 10 metallicities: 2, 1, 1/2, 1/5, 1/10, 1/30, 1/50, 1/100, 1/1000,
and 0 times the solar metal composition. So that the grid is useful for studies of typical environments of massive stars: the Galactic center, the Magellanic
Clouds, blue compact dwarf galaxies like I Zw-18, and galaxies at high redshifts.
For model atmospheres, the grid is available at the TLUSTY Web site\footnote{http://tlusty.oca.eu/Tlusty2002/tlusty-frames-BS06.html}.
Considering the typical characteristics of the star of NGC\,1850, we adopt the stellar atmospheric models with Z=0.008 and microturbulent velocity of 2 Km/s. 
In fact, recent studies have determined the metallicity of the cluster, \cite{antony22} and  \cite{song21} found a  value [M/H] = -0.31 (Z $\sim$ 0.008), \cite{kamann23}  found -0.33. 
The standard solar abundances adopted are by \cite{asplund05}.

The model atmospheres explicitly include, and allow for departures from LTE, 46 ions of H, He, C, N, O, Ne, Mg, Al, Si, S,and Fe, and about 53000 individual atomic levels grouped into 1127 superlevels. For more details see \cite{syns54,Lanz07} and reference therein.
The triplet lines of He at 4922 \AA\ are treated using special line broadening tables,  in accordance with  \cite{barnard74} \cite{shamey69}.

To have He enhanced  synthetic spectra is not advantageous to construct a model from scratch, so we used the existing model from the grid to interpolate the values for the required helium abundance.
A conservative estimate is to change the chemical abundance no more than 0.2 dex \citep{tlusty3}.
Since the difference in helium abundance  between the populations could 
ranges from less than 0.01 to more than 0.2  in He mass fraction (Y) (e.g. \citealt{dantona04}, \citealt{milone15},\citealt{gratton19}, \citealt{milone22} and reference therein), for each of these combinations in $T_{eff}$-$\log g$, models with a helium abundance of logN(He)/N(H)= 0.00, +0.04, +0.1, and +0.2 dex were computed, corresponding with a helium abundance in mass fraction of Y= 0.25, 0.27, 0.30 and 0.35.

\begin{figure}
	\includegraphics[width=\columnwidth]{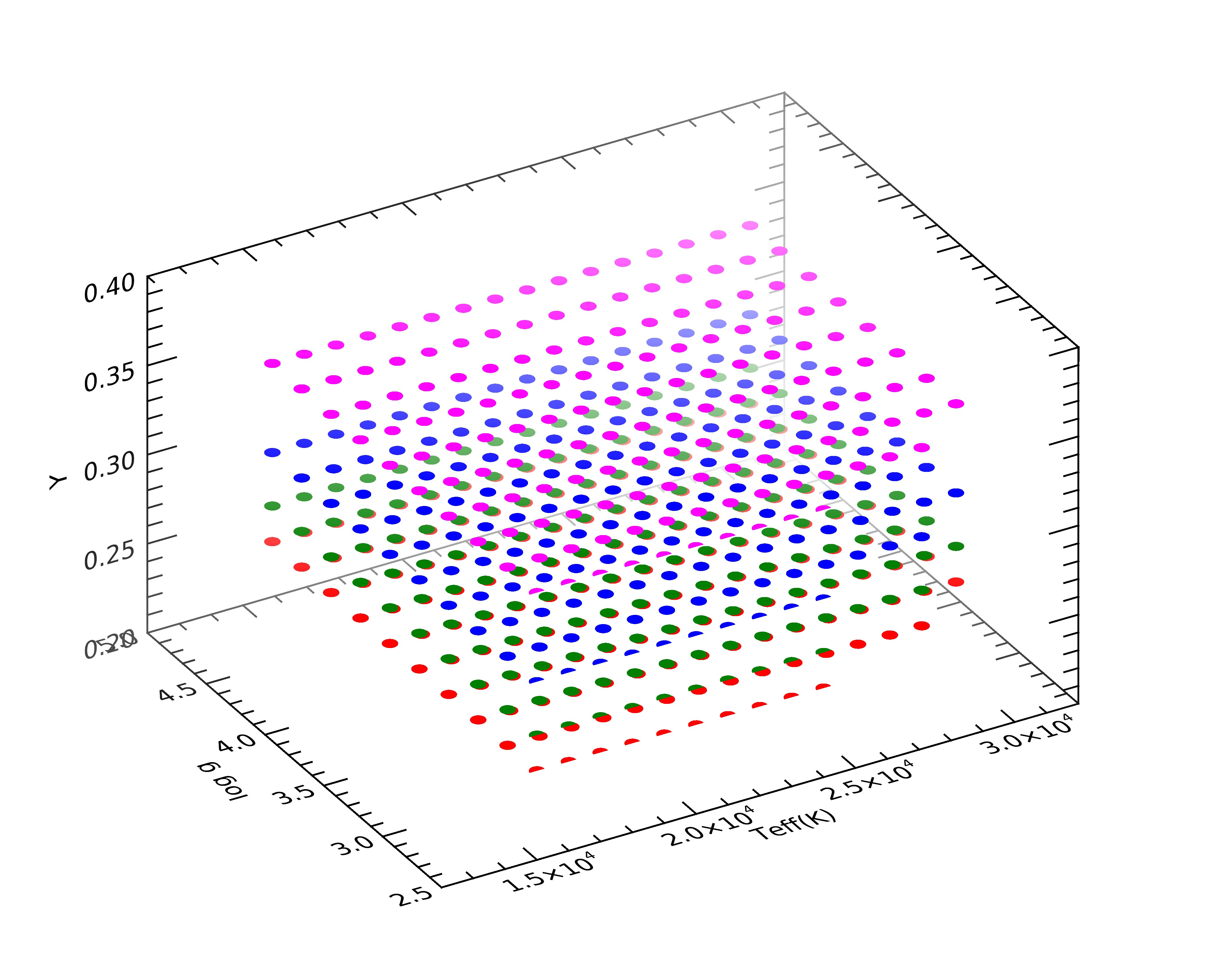}
    \caption{Grid points in the $\log g$ vs. $T_{eff}$ vs $Y$ plan. Red, green, blue and magenta points represent different He mass fraction Y, 0.25, 0.27, 0.30 and 0.35 respectively. For a colour version of the figure, see the electronic version of the paper.
	}
    \label{th}
\end{figure}

The whole grid  points are shown in Fig. \ref{th}, in the $\log g$ vs. $T_{eff}$ vs $Y$ plans.

\begin{figure}
	\includegraphics[width=\columnwidth]{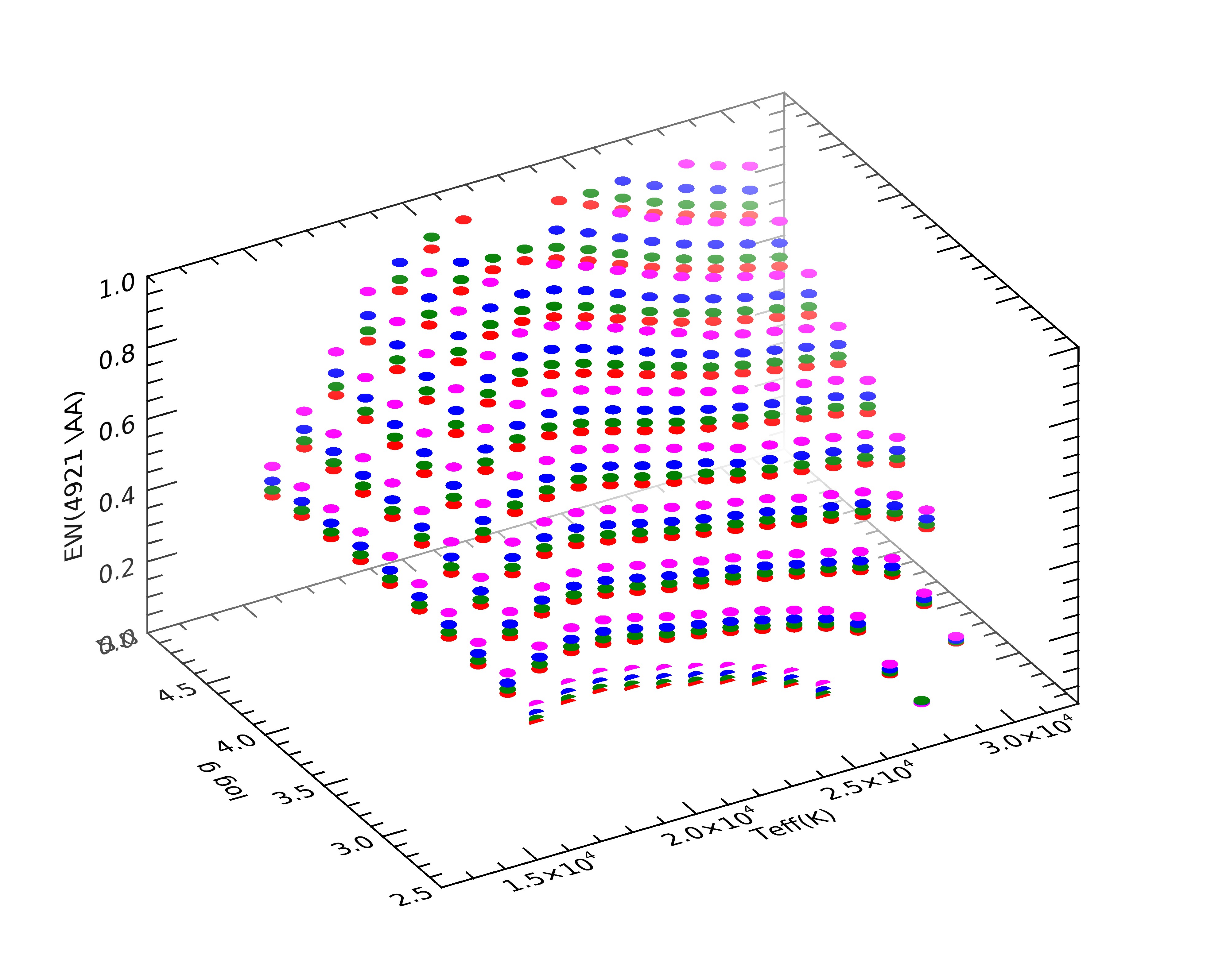}
    \caption{Equivalent width of the 4922 \AA\ line in the $\log g$ vs. $T_{eff}$ vs $Y$ plan. The symbols are as in Fig. 1. 
	}
    \label{4922}
\end{figure}

To obtain a synthetic spectrum that is directly comparable to observations,
one has to convolve the net emergent flux  with the instrumental profile of the spectrograph that
produced an observed spectrum to be analyzed. To this end, 
the synthetic spectra were degraded at the MUSE resolution and normalized  with ROTIN code \citep{,hubeny17}.
For each spectrum, we have determined the equivalent width (EW) of 4 He absorption lines, at 4922 \AA, 5015 \AA, 6678  \AA\, and 7065 \AA.
We omitted  the most intensive He line at 5876 \AA $ $ because it  is not available in the MUSE spectra, since it falls in the portion of the spectrum masked.
We obtained these values numerically integrating the  normalized spectrum  considering  10 \AA-wide region on both sides of the He I lines.
As example, in Fig. \ref{4922}   we show the equivalent width of the 4922 \AA\ line, calculated from the synthetic spectra.\\
As shown in Fig. \ref{4922} the behaviour of the intensity of the EW is in agreement with the theory. As expected:
\begin{itemize}
\item The EWs increases with the He abundance;
\item Since lines of neutral helium first show up in the O-type stars, strengthen through the O-type stars, come to a maximum at a spectral type B2 on the main sequence , and the weaken toward later (cooler) type, the behaviour of the EWs has a  maximum at 16000- 22000 K, and decreases for cooler  and hotter temperatures;
\item The intensity of the lines increases with the surface gravity.
\end{itemize}

Moreover, for values of $T_{eff}$ less than 
17000 K the strengths of this line are quite sensitive to
temperature but relatively insensitive to gravity. For $T_{eff}$ between   18000 K and 22000 K the abundance is dependent on $\log g$ more than $T_{eff}$ At still higher temperatures, both $T_{eff}$ and $\log g$ must be accurately determined in order to derive the abundance of He from the measured equivalent widths \citep{wolff85}.


To compare the  theoretical equivalent width ($EW_{th}$) with the observed ones ($EW_{obs}$), we have interpolated  through the non-LTE grid of synthetic EWs as a function of $T_{eff}$ and $\log g$.

\section{Data Analysis}

\label{sec2}

Spectra were extracted  from the final MUSE data cubes using specifically developed software  by \cite{antony22} (for the details see Par 2 of the paper). 
The extraction was performed on each individual data cube, included the combined cube (cube$_{sum}$ hereinafter).
We have extracted the   spectra  of the same 1167 stars from each cube.
Residuals from the strong nebula emission lines were still visible in some of the extracted spectra,  but they do not affect the He lines.
The region analyzed is not heavily influenced by nebularity. In fact, analyzing the spectra of background 
 in the regions nearby our selected stars,
we have verified the absence of the He lines in emission.
To identify the MUSE targets and to determine the effective temperature ($T_{eff}$) and gravity ($\log g$) of them we used the archival  \textit{Hubble Space Telescope} (HST) photometry of NGC\,1850 taken with
the Wide Filed Planetry Camera 2 (WFPC2) on March 4, 1994, during the  program  \#5559 (PI:Gilmozzi).
The data, reduced using $DOLPHOT$ software \citep{dolphin}, include the images taken in three filters $F170W$, $F439W$ and $F569W$ and they  have been already  presented in \cite{gilmozzi94} and \cite{antony22}.

\begin{figure}
	\includegraphics[width=\columnwidth]{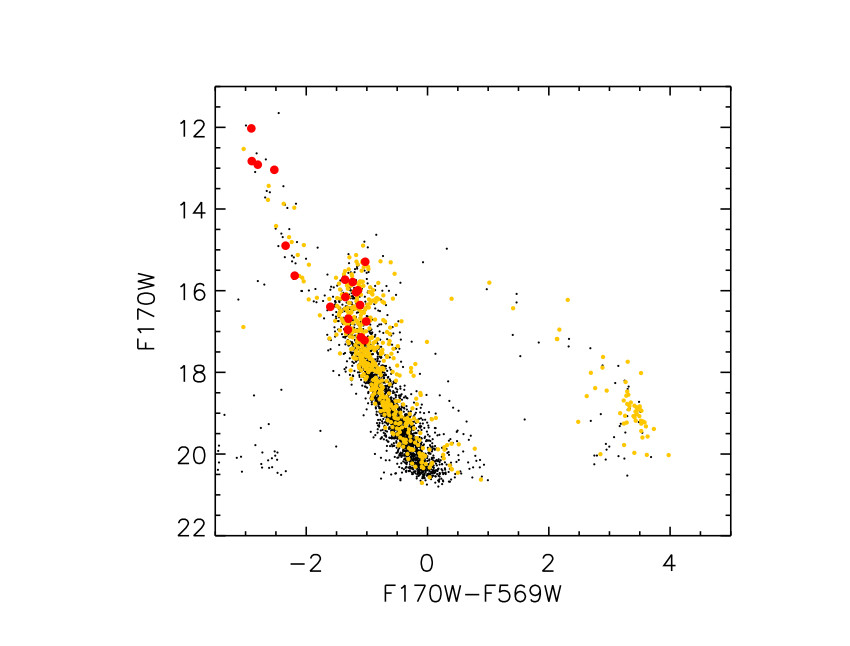}
    \caption{Color-magnitude diagram of the stars in the NGC 1850 area. Black points are the HST data, yellow points are the stars extracted from MUSE data cube, and the red point are the B stars analyzed here, of which we have determined He abundance.
	}
    \label{cmd}
\end{figure}

In Fig. \ref{cmd} we show the color-magnitude diagram  in $F170W$ and $F569W$ bands. The black points are the HST data, the yellow ones represent the MUSE targets extracted, the red ones are the B stars selected to  estimate the He abundance (see Sec. \ref{selection}).
The MUSE sample covers a wide range of color  and magnitude,  it is possible recognize the blue supergiants of NGC\,1850 B ($F170W$ - $F569W$< -2) the MS, RG and blue loop stars of NGC\,1850 A ($F170W$ - $F569W$ > -2 ). A contamination from LMC stars is also present ($F439W$ - $F569W$ > 0.3 and  $F569W$ > 17, the magnitude in the band $F170W$ of these targets is not present in the catalog).
 Relevant quantities of the stars of our  sample  selected (see sect. 4.1) are reported in  table 1, they were determined by comparing them with appropriate theoretical isochrones  ( see \cite{antony22} for more details about this technique). We used \cite{marigo08}'s set of solar-scaled isochrones with a metallicity of Z = 0.008 and ages of 15 Myr, 90 Myr, and 10G yr.
     Stars with the color  $F170W-F569W$ $<-2$ or stars within 5" from the center of the cluster NGC\,1850 B  to the youngest one,  stars belong to NGC\,1850 A are fitted with the isochrone at 90 Myr, and  stars of LMC have been associated with the  oldest  isochrone.
The distance modulus $(m-M)_0$= 18.50 \citep{niederhofer15} and the reddening $E(B-V)$ =0.13 \citep{gorski} have been adopted.
Once  the $T_{eff}$ and $\log g$ have been determined,  the radial velocities of the targets stars have been derived  by cross-correlating the stellar spectra with the appropriate and  
previously calculated  synthetic spectra.
Cross-correlation was performed using  $fxcorr$ task  from $IRAF$, considering the available He lines.
As noted by \cite{antony22}, the accuracy of the radial velocity varies greatly with the color of the targets, for this reason the values found for the B stars in this work are  slightly different from the average velocity found by \cite{antony22}, \cite{fischer93}  (251.4 $\pm$ 2 Km/s) and \cite{song21} (248.9 $\pm$ 2.5 Km/s), who have derived the radial velocity from a sample of RSG stars which are much cooler than the stars used here.
Finally, we shifted the spectra to the rest frame.

\begin{table}
	\centering
	\caption{Relevant quantities of the stars of our  selected sample.}
	\label{teff}
	\begin{tabular}{ccccccc} 
		\hline
		Id & Ra & Dec& Teff & Log g & v$_{rad}$&$\sigma_{v_{rad}}$\\
               &    &   & (K)    &  (dex)   & ($Km/s$)  & ($Km/s$)           \\
		\hline
            \multicolumn{6}{c}{\textbf{NGC\,1850 A}}\\
            \hline
		222  & 77.2163239 & -68.7669383 & 15200 & 4.16 & 270 & 8\\
		377  & 77.2080954 & -68.7678015 & 15000 & 4.10 & 266 &  8\\
            415	 & 77.206     & -68.761     & 20300 & 4.27 & 270 &  7\\
            511  & 77.2018423 & -68.7656164 & 16100 & 3.85 & 285 & 5\\
            564  & 77.1990042 & -68.7531787 & 17900 & 3.98 & 267 &  10\\
            727  & 77.1924089 & -68.7728569 & 16600 & 4.15 & 261 &  17\\
            753  & 77.1915224 & -68.7681341 & 15700 & 4.16 & 255 & 10 \\
            907  & 77.185     & -68.774     & 16800 & 3.92 & 275 & 10\\
            952  & 77.1835724 & -68.7753472 & 16600 & 4.10 & 259 &	9\\
            971  & 77.1830299 & -68.770169  & 15800 & 3.98 & 280 &	11\\
            1036 & 77.1795715 & -68.7560604 & 15100 & 3.34 & 266 & 9\\ 
            1069 & 77.1767874 & -68.761876  & 16000 & 3.85 & 253 & 8\\  
            1129 & 77.1703261 & -68.7598117 & 16800 & 3.77 & 235 & 7\\
           
		\hline
             \multicolumn{6}{c}{\textbf{NGC\,1850 B}}\\
             \hline
             605  & 77.1974417 & -68.7680385 & 21400 & 4.27 & 254 &6 \\
            1020 & 77.1805277 & -68.7682749 & 25100 & 3.94 & 261 &	10\\
            1077 & 77.1760504 & -68.7641924 & 23000 & 4.20 & 257 & 9\\
            1146 & 77.1673117 & -68.7613749 & 29900 & 3.83 & 255 &	10\\
            1153 & 77.1655824 & -68.7616053 & 30000 & 3.62 & 262 &  8\\
            1160 & 77.1639984 & -68.7606044 & 28500 & 3.83 & 257 & 10\\
            1161 & 77.1635335 & -68.7617016 & 25800 & 3.83 & 269 & 8\\
            \hline
	\end{tabular}
\end{table}

\subsection{Selection criteria for our sample of B stars in NGC 1850 A and NGC 1850 B}
\label{selection}

On the basis of the 1167 stars in our sample of MUSE spectra, we selected the stars with a  $T_{eff}$ between 15000 K and 30000 K,
the surface gravity $\log g$ between 2.5 and 4.75, values of the theoretical grid (see Sec. \ref{sec:syns}). We also selected the ones with at least two available measurements of magnitude in the bands $F170W$, $F439W$ and $F569$, this is to secure a fairly reliable determination of the $T_{eff}$ for the stars included in the selected sample. In this way, we end up with a sample reduced to 232 targets.

Moreover, we excluded the  Be stars  to minimize  uncertainties due to rotation and complexity in the spectra analysis,  identified spectroscopically through the $H\alpha$ emission.
We found a fraction of Be stars  a little bit higher ($\sim$ 65\% of the sample) than  one found  in NGC\,1850 by \cite{bastian17} ($\sim$ 20-50\%) and in other young massive cluster \cite{milone18} (40-55 \%).

We have also excluded the spectra showing stellar activity, the ones in which the He lines are too faint to be fitted or  the signal to noise (S/N) is lower than 30.
In conclusion, our selected simple is composed by  20 B stars, all these stars are highlighted  as red point in Fig. \ref{cmd}.
This sample includes B stars of NGC\,1850 A and B as reported in table \ref{teff}.
For simplicity, we consider stars of NGC\,1850 B each target associated to the youngest isochrone, even if the distance from the center of the sub-cluster  is more than 5''.


\subsection{He abundance}

In this section, we make use of the spectral data secured by MUSE observations to evaluate the helium abundances in the B stars of NGC 1850 A and B, and to investigate the presence of a possible spread of this quantity in our sample.

\subsubsection{\textbf{EWs determination}}

To estimate the He abundance we determine the equivalent width  of the lines at 4922 \AA\, 5015 \AA\, 6678 \AA\, and 7065 \AA\ from the spectra observed, these values are then compared with the theoretical ones.

For the middle B spectral types, the interpretation of the He I at 4921.9 \AA\, is complicated by the blends with Si II  at $\lambda$ 4921.7 \AA\ and with lines of Fe II and S II at $\lambda$ 4923.9 \AA.
While for B0 and B1, O II lines at $\lambda$ 4924.6 \AA\ perturbs the red wing of He I $\lambda$ 4921.9 \AA\ .
Moreover, middle and late B spectra are affected by S II at $\lambda$ 5014.0  \AA\ blending with He I at $\lambda$ 5015.7  \AA\ \citep{leck71}.
 All these issues are well known, nevertheless we recall them to remind that the evaluation of He abundances in this range of wavelength has to be regarded with special care. For this reason, our results will be used to enlighten qualitative new discoveries which will lead to further work to be quantitatively confirmed. 


In the present work, each He line was fitted individually to derive the EWs.
As a first step, the continuum 
was estimated via a linear fit of the flux obtained in two narrow range (10 \AA) of wavelength near the wings of each He absorbing line we intend to evaluate. This local continuum has been used to normalize the part of the spectra where it is present the He line.

\begin{figure}
	\includegraphics[width=\columnwidth]{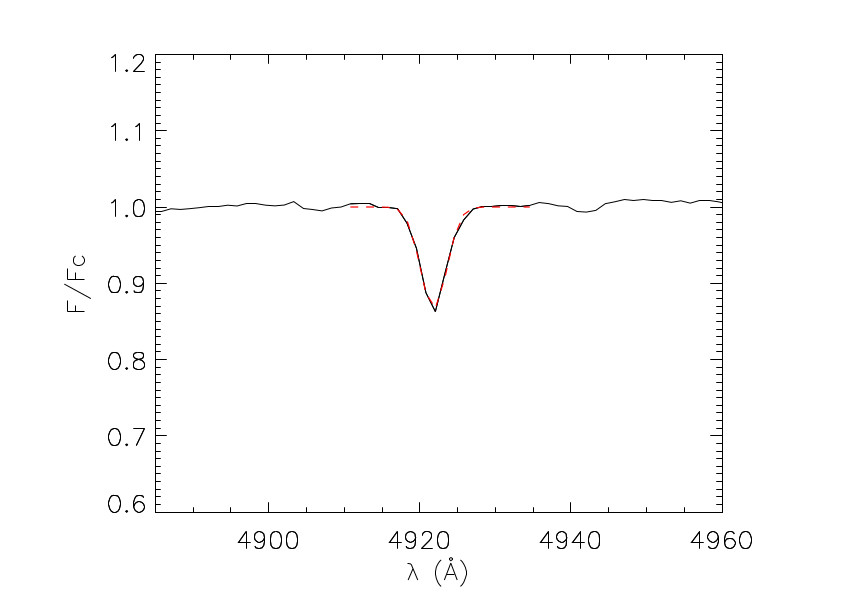}
    \caption{Best fit of the He line at 4922 \AA\ for the star \#1146. The black line is part of the spectrum extracted near the line, the red dashed line is the best fit. 
	}
    \label{fit}
\end{figure}

 The previous step make possible to evaluate the EW of each He line by performing a Gauss fit. As an example of the quite good quality of this step,  Fig. \ref{fit} shows  the best fit  (red dashed line) obtained for the star \#1146 at 4922 \AA.

\begin{figure*}
	\includegraphics[width=\textwidth]{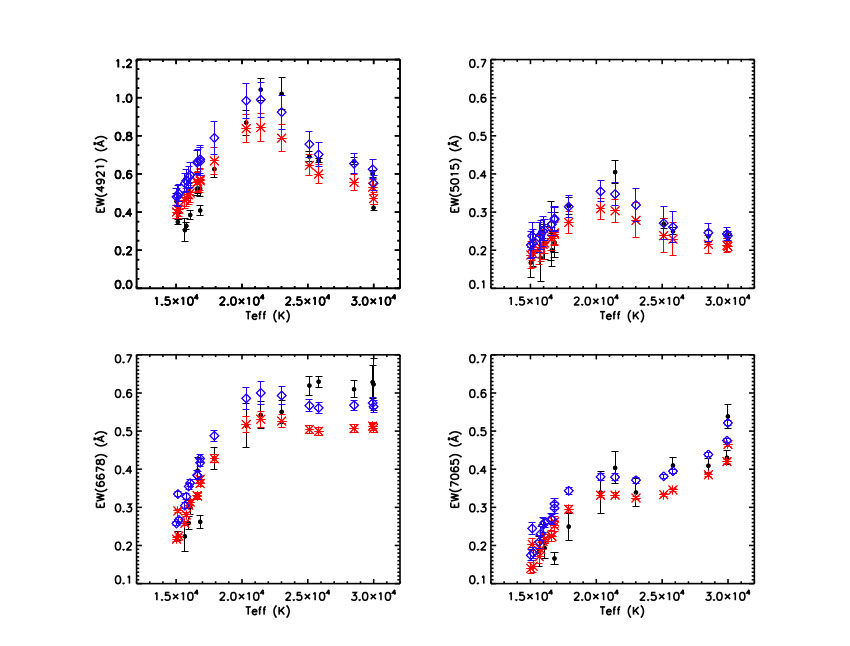}
    \caption{Equivalent width (\AA) for each line  obtained from MUSE spectra (black dots) and from the interpolation of the  theoretical $EW_s$ with different helium mass fraction: red  asterisk and blue diamonds for models with Y=0.25 and  Y=0.35 respectively.}
    \label{ewew}
\end{figure*}

Unfortunately, the strongest line at 5876 \AA\  has not been taken into account
because the section of He line near
the sodium doublet has been masked to avoid the strong emission due to the laser guide.
It is relevant to report that none of these stars, but \#971, exhibit a significant variation in radial velocity  or in the EW values among the six epochs. 
This means that the stars of our sample can be considered single stars, or their companion is a low-mass star that does not significantly affect the radial systemic velocity. 
 Since the S/N in the $cube_{sum}$ is systematically higher 
 than the S/N found in each single $cube$, and the values of the  mean EWs of each lines of the single cubes  are in agreement,  within the error,  to the EW values of the cube$_{sum}$ ,  we decide to continue the analysis only with the values estimated from the last one. In table \ref{tablehe}, we present the EWs of the He lines at 4922 \AA, 5015 \AA, 6678 \AA, and 7065 \AA. 
Empty values in this table refer to features for which the best fit was not obtained, in most cases because the observed line was too weak or not detectable in the noise.
In Fig. \ref{ewew} we show the $EW_{obs}$ (black points) for each star in each He I lines as function of temperature in comparison with the ones calculated with different helium mass fraction, Y=0.25 (red asterisks) and Y=0.35 (blue diamonds). 
 For sake of clearness, the plot does not include the $EW_{th}$ relative to Y=0.27 and Y=0.30. This figure shows a quite good agreement between the trend of the $EW_{th}$ and the one of $EW_{obs}$ as a function of $T_{eff}$. This result supports our previous assumption about the NLTE, which appears to be the most appropriate to analyse the helium abundance in B stars.

\begin{table*}
\scriptsize
\caption{ Equivalent width (in \AA) obtained from our analysis for each line and the derived mean Helium abundance ($Y$). 
}
\label{tablehe}
\begin{tabular}{ccccccc}
\hline
Star&$EW$&$EW$&$EW$&$EW$&  $<Y>$ & $\sigma$(Y) \\ 
  & ($\lambda$4921.9 \AA) & ($\lambda$5015.7 \AA)  &  ($\lambda$6678.2 \AA) & ($\lambda$7065.7 \AA) &   &     \\ 
\hline
 \multicolumn{6}{c}{\textbf{NGC\,1850 A}}\\
 \hline
 222 &   0.40 $\pm$   0.04  &  0.19  $\pm$  0.04  &  ...                  &  ...               & 0.24 & 0.02 \\
 377 &   0.46  $\pm$  0.04  &  0.17  $\pm$  0.04  &  ...                  &  ...               & 0.25 & 0.10 \\
 415 &   0.87  $\pm$  0.07  &  ...                 &  0.52  $\pm$   0.60  &  0.34 $\pm$  0.06 & 0.26 & 0.01 \\
 511 &   0.39  $\pm$  0.03  &  0.22  $\pm$  0.04  &  0.30  $\pm$   0.02  &  0.19 $\pm$  0.03 & 0.20 & 0.05 \\
 564 &   0.63  $\pm$  0.04  &  0.32  $\pm$  0.02  &  0.43  $\pm$   0.03  &  0.25 $\pm$  0.04 & 0.24 & 0.08 \\
 727 &   0.52 $\pm$   0.02  &  0.20  $\pm$  0.04  &  0.40  $\pm$   0.03  &  ...               & 0.25 & 0.11 \\
 753 &   0.31 $\pm$   0.06  &  ...                 &  0.22  $\pm$   0.04  &  0.18 $\pm$  0.04 & 0.18 & 0.10 \\
 907 &   0.41 $\pm$   0.03  &  0.22  $\pm$  0.03  &  0.26  $\pm$   0.02  &  0.17 $\pm$  0.02 & 0.10 & 0.06 \\
 952 &   0.52 $\pm$   0.03  &  0.25  $\pm$  0.08  &  ...                  &  ...               & 0.26 & 0.06 \\
 971 &   0.33 $\pm$   0.02  &  0.18  $\pm$ 0.06   & ...                   &  ...               & 0.13 & 0.06 \\
 1036 &   0.35 $\pm$   0.02  &  0.20  $\pm$ 0.02   &  0.23  $\pm$   0.01  &  ...               & 0.16 & 0.06 \\
 1069 &   0.51 $\pm$   0.04  &  0.25  $\pm$  0.02  &  0.26  $\pm$   0.02  &  0.23  $\pm$ 0.02 & 0.27 & 0.08 \\
 1129 &   0.52 $\pm$   0.04  &  0.24  $\pm$  0.06  &  ...                  &  ...               & 0.22 & 0.02 \\
 
\hline
\multicolumn{6}{c}{\textbf{NGC\,1850 B}}\\
 \hline
  605 &   1.04  $\pm$  0.06  &  0.41  $\pm$  0.03  &  0.54 $\pm$   0.04  &  0.40 $\pm$  0.04 & 0.38 & 0.09 \\
  1020 &   0.69 $\pm$   0.03  &  0.27  $\pm$ 0.01   &  0.62  $\pm$   0.03  &  ...               & 0.36 & 0.07 \\
  1077 &   1.02  $\pm$  0.08  &  ...                 &  0.55  $\pm$   0.03  &  0.34  $\pm$ 0.04 & 0.33 & 0.08 \\
 1146 &   0.60 $\pm$  0.02  &  0.24  $\pm$  0.01  &  0.63  $\pm$   0.04  &  0.43  $\pm$ 0.02 & 0.34 & 0.07 \\
 1153 &   0.42 $\pm$   0.02  &  0.23  $\pm$  0.01  &  0.62  $\pm$   0.07  &  0.54  $\pm$ 0.03 & 0.34 & 0.11 \\
 1160 &   0.67 $\pm$   0.02  &  0.24  $\pm$  0.02  &  0.61  $\pm$   0.02  &  0.41  $\pm$ 0.02 & 0.35 & 0.05 \\
 1161 &   0.67 $\pm$   0.01  &  0.25  $\pm$  0.01  &  0.63  $\pm$   0.015 &  0.41  $\pm$ 0.02 & 0.37 & 0.07 \\
 \hline
\end{tabular}
\end{table*}

\subsubsection{\textbf{He mass fraction evaluation}}

\begin{figure}
	\includegraphics[width=\columnwidth]{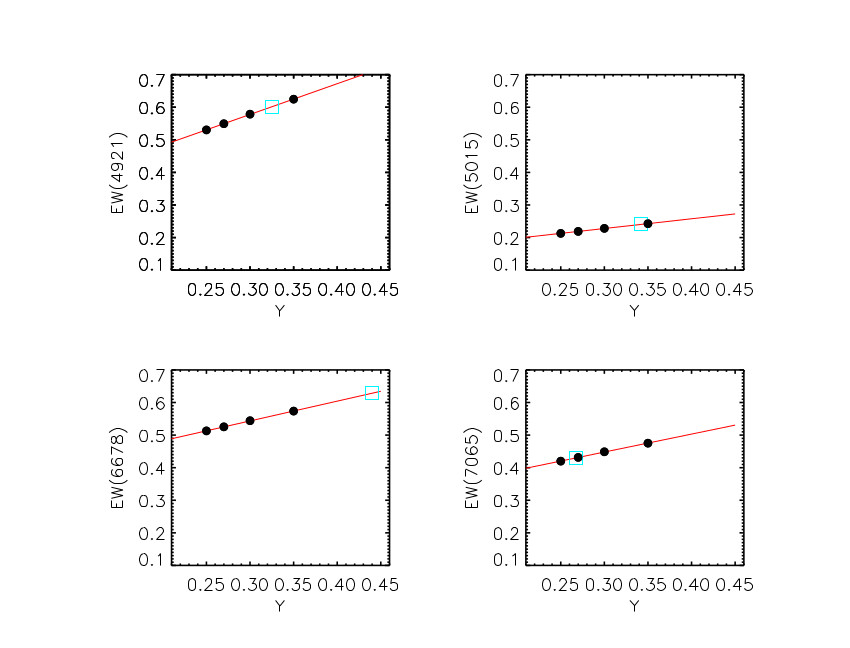}
    \caption{Values of the $EW_{obs}$ (cyan square) and the $EW_{th}$ (black points) corresponding at Y=0.25, 0.27, 0.30 and 0.35,  for the  star \#1146 . 
  The red lines represent the linear fit of the synthetic data.}
    \label{ew_int}
\end{figure}

An evaluation of the He mass fraction (Y) from the $EW_{obs}$ of the He lines  can be obtained by using the synthetic computed for these lines. To this purpose we used a non-LTE grid of  $EW_{th}$ covering a wide range of effective temperature and gravity.
The precise values  of Y are then obtained by accurate interpolations through the non-LTE grid of $EW_{th}$ for each line as a function of gravity, temperature and helium, when the measured quantities ($EW_{obs}$) are within the border values of the grid. In case the value is out of the available grid, we obtain a fair evaluation of the Y by using a linear extrapolation.  
We show an example of both the cases in Fig. \ref{ew_int} for the star \#1146.  
The cyan point is the measured equivalent width from the MUSE spectrum while the black points are the values of the synthetic equivalent widths  corresponding at Y=0.25, 0.27, 0.30 and 0.35. The red lines represent the linear fit of the synthetic data.\\
Finally, we derived the He abundance by computing the mean of the abundances found for each available line, the results are reported in column 6 of the table \ref{tablehe} with the standard deviation associated (column 7).
{It is evident from  of our analysis that we can divide our sample in 3 groups, $He-normal$ star, $He-weak$ stars and the $He-enhanced$ stars, or $He-rich$.
Even if the typical He  mass fraction value for LMC stars is Y=0.25, considering the uncertainties on the determination of the He abundance, we call  $He-normal$  stars the targets with Y between 0.24 and 0.26.  
The second group is populated by targets showing  Y<0.24, while the third is the sample collecting the stars with Y>0.26.}

 We plot  the targets  of the three groups in the CMD with different symbols (fig. \ref{rot}),  it is evident  that they  arrange in specific part of the CMD:  
 \begin{itemize}
 
 \item "$He$-$rich$" stars, with the exception the target  \#1069, are located in the region of the  the MS associated to the  isochrone at 15 Myr, with which we identify the NGC\,1850 B cluster;

\item Most of the "$He$-$normal$" stars (green diamonds) have a F170W magnitude larger than -2 mag, they are present only in the  MS that collect the stars belonging to  NGC\,1850\,A.

\item The "$He$-$weak$" stars (light blue squares) seems to populate the reddest part of the MS (F170W lower than -2 mag).

\end{itemize}

According to \citep{osawa,garrison94} it is not surprising to find  He-weak stars, because they constitute a sub-class of B-type stars .
The He-weak stars have effective temperatures of typical B3-B5 stars, as in our sub-sample 
 and the weak helium lines in the spectra of certain stars cause their spectral types to be inaccurately determined, leading to a discrepancy between their spectral type and their apparent color. Moreover,  there is evidence that their spectra show variations in the metal lines and perhaps helium lines \citep{molnar72}, as  could be the case of the target \#197 of our sample.

The He-rich targets (red points), are in the hot and bright part of the CMD, i.e. where the younger stars are located and in the region where we identify most of the stars of NGC\,1850 B. 
On average the Y value of these stars is 0.35 $\pm$ 0.02 . 

 We note that  4 stars of this sub-sample are located at a distance from the center of NGC\,1850 B less than 5'', namely \#1146,  \#1153, \#1160, \#1161. The other three are at larger apparent distance from the center, however a quick check by eyes shows they seem to lie in the edge between the two sub-clusters.


Quite interestingly we find that some of the He-rich stars are in common with the weak-Oxygen targets found by \cite{antony22}. If confirmed, this could be extremely interesting just because this kind of anti-correlations  is quite typical of the multi-populations in Galactic GCs, in which the second generations of stars are rich in Helium and poor in Oxygen.

However, we remark that the two samples of stars (He-normal and He-
rich) do not overlap in effective temperature, thus hampering 
a direct comparison and possibly leading to unknown sources of systematic error.

\subsection{Uncertainties}

The internal uncertainties on our He abundances include errors due  to the determination of  the atmospheric parameters.
We evaluated the effective temperature and gravity of the B stars by comparing  the theoretical models having the same  color and magnitude of stars in our sample.
The uncertainties on these quantities  are dominated by the uncertainties on the magnitudes and distance.
The photometric error of our data is about 0.02 mag \citep{gilmozzi94}.
At the typical temperature of our sample, this translates to a temperature uncertainty of $\sim$ 1000 K and in gravity of $\sim$ 0.01 dex.
To estimate the internal uncertainties  associated with the  helium contents, we re-computed the abundance for a set of B stars varying the initial atmospheric parameters by (adding/subtracting) a quantity equal to their expected errors.
By summing in quadrature the two contributions, we estimate the total uncertainty in Y as large as 0.04, being dominated by errors from temperature. 
 Except for few stars, this evaluation of uncertainty appears smaller than  
the $\sigma$ values  estimated along the procedure of measuring the He abundance (see Tab \ref{tablehe}).

A systematic error on the atmospheric parameters estimation could be due to the use of different grid models respect TLUSTY ones to convert the theoretical isochrones to the WFPC2 photometric system.
In fact, Marigo and collaborators  used  the transformations primarily based  on the  LTE ATLAS9 \citep{castelli} synthetic atmospheric models.

On this matter, we note that the LTE ATLAS9 and non-LTE TLUSTY model atmospheres both assume the same microturbulence, i.e. 2 Km/s\footnote{https://wwwuser.oats.inaf.it/fiorella.castelli/grids.html.}
Furthermore,
\cite{przy11}  tested the LTE ATLAS9 and NLTE TLUSTY model atmospheres, concentrating their work on the  early-type stars of effective temperature between 15000 and 35000 K.
They found that  the temperature structure of the two models are in agreement, the differences 
are at most  1-2\%, and they are  even smaller  for low metallicity.
As a result  of this work, they concluded that such a  small difference is irrelevant  for the stellar parameter determination.
Moreover, \cite{Lanz07}  compared  the same models atmospheres  for   B stars.
The authors 
find negligible differences between the continuum  of the two spectra sets.
The strongest difference is in the  near ultraviolet
range, where the LTE fluxes  are about  10\% higher then the NLTE predictions.  
This leads in a difference in the estimation of the effective temperature using HST broad bands lower than the uncertainties 
reported in our paper. 
We 
conclude that  the use of the ATLAS9 models instead of TLUSTY ones does not affect the results of this work.

The possible systematic errors on the atmospheric parameters due to the uncertainties in the adopted reddening have been also evaluated.
To this purpose, we re-computed the temperature and gravity of each star by  varying the initial reddening in an artificial way, i.e. by adding/subtracting to the original reddening of a quantity equal to their expected errors (0.015) \citep{gorski}.
   As a result, we have obtained mean differences in effective temperature about 200 K and in log g $\simeq$0.01 dex, both values well within the uncertainties considered in this work.

Another source of systematic uncertainty could affect the He abundance is the microturbulence. 
Microturbulence is sensitive to the stellar atmospheric parameter, especially surface gravity \citep{hunter07}.\\
 By analyzing spectra of 102 B stars,  \cite{lyub04} found that stars with mass between $\sim$ 4 and 7 $M_\odot$ have a microturbulence velocity ($V_t$) between 0 and 5 Km/s, with a average about 1.7 Km/s. While stars with mass in the range of 7 - 11 $M_\odot$ have the same spread and average value of microturbulence of the previous case until their  relative age $t/t_{MS}$ is less then 0.8. After this value the microturbulence velocity could arise up to  $\sim$ 11 Km/s, with an average of about 7 Km/s.
For the most massive stars (12-18 $M_\odot$)  the microturbulence velocity  spreads between 4 an 23 Km/s, and it depends on the relative age $t/t_{MS}$.\\

Recently, \cite{liu2022} found an empirical relationship between the microturbulence and surface gravity for the B-type stars:
\begin{equation*}
    V_t=-3.97*(\log g)^2 + 17.85*(\log g) -2.52
\end{equation*}


The choice of 2 Km/s for our theoretical models is in agreements with the quoted studies, but for the hottest and massive ones \#1146,\#1153,\#1160  \#11161, for which the more accurate $V_t$ should be about  7 Km/s.
 We believe that microturbulence may play a role in constraining the He abundance in young hot stars as the ones we studies here. Nevertheless,   it does not seem that uncertainties of microturbulence can have an impact so sharp to generate the He-enhancement we found here.

\subsection{Rotation}

Since  stellar rotation contributes to the broadening of line profiles in the stellar spectra,  in this paragraph we discuss its possible role in the EWs values  of the He lines in the NGC1850 B stars spectra. 
The aim is to understand if the increase of the $EW_{obs}$ associated to  the He line profiles is due to the rotation or to the He-enhanced.
 Since  the resolution of MUSE spectra  is not high enough to  analyze the issue in detail, we 
provide a qualitative analysis.

\begin{figure}
	\includegraphics[width=\columnwidth]{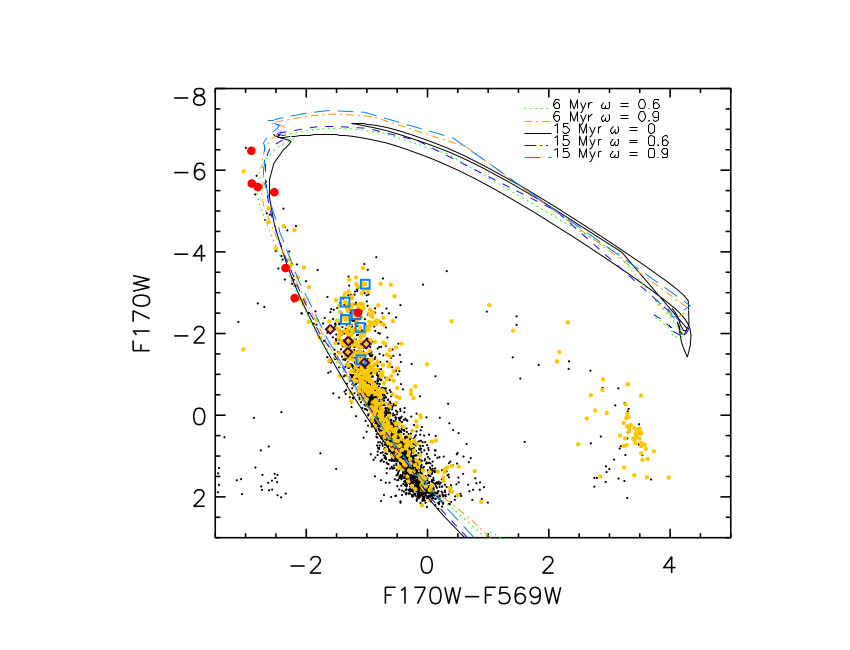}
\caption{(F170W-F569W,F170W) CMD of NGC 1850 with superimposed isochrones with different rotation rate  and age from \protect\cite{marigo08} and 
\protect\cite{nguyen22}. Isochrones are colour-coded, in terms of their  $\Omega/\Omega_{cr}$ and age, as follows:$\Omega/\Omega_{cr}$=0 and age  15 Myr black line; $\Omega/\Omega_{cr}$=0.6 at 15 Myr purple dashed line; $\Omega/\Omega_{cr}$=0.9 at 15 Myr dark green long dashed line;$\Omega/\Omega_{cr}$=0.6 at 6 Myr  green dotted line, $\Omega/\Omega_{cr}$=0.9 at 6 Myr  cyan dash-dotted line.
    Targets associated with stars He-enhanced  are marked by red dots.}

    \label{rot}
\end{figure}

To find a plausible rotation velocity of B stars of NGC1850 B, we have compared the CMD   with the isochrones of rotating models obtained from PARSEC version 2.0 \citep{nguyen22}.
As an input indication for the age of these models, we took into account the work by \cite{antony22} and \cite{fischer93} that fit the CMD of NGC\,1850 B with a non-rotating isochrone of 15 Myr and 6 Myr respectively.
Thus, we compare the HST photometry to 5 isochrones at 6 Myr and 15 Myr computed with different rotation rate $\omega=\Omega/\Omega_{cr}$ 0,0.6 and 0.9,  where $\Omega$ is the angular velocity and  $\Omega_{cr}$ is the breakup velocity, that is the angular velocity at which the centrifugal force is equal to the effective gravity at the equator.
As shown in Fig. \ref{rot},  the CMD of NGC 1850 B
is pretty well fitted by all  the isochrones taken into account, both rotating and non-rotating.
By adopting these theoretical models, 
the stars having F170W $\sim$ -6  are reproduced by isochrones of 6 Myr and 15 Myr and $\omega$ =  0.6 disclosing equatorial velocity $v_{eq}$ $\simeq$ 230 - 250 Km/s.
On the other hand, if the isochrones with $\omega$ = 0.9 are adopted, an equatorial velocity of $v_{eq}$ $\simeq$  410 Km/s is found.


Since  velocities of 230-250 Km/s are derived for some stars of NGC 1850 B in the analysis by \cite{kamann23} and 
 by \cite{antony22},  we decided to compute the theoretical spectra with $ROTIN$ by assuming $Y = 0.25$ and a rotation velocity of 250 Km/s. 
Also in this case the synthetic spectra are degraded to the MUSE resolution and normalized.
We considered the same  set of $T_{eff}$ and $\log g$ of the grid {\it BSTAR2006}.
The same procedure explained in sec. \ref{sec:syns} has been used  to derive the $EW_{th}$ of these models for each He lines.
Since we are taken into account  the  case of $\Omega/\Omega_c$ =0.6, the effect of the rotation on the atmospheric parameters can be neglected \citep{fremat05}.

As an indicator of the quality of the derived $EW_{th}$, we determined the mean value of the difference $\Delta_{EWrot}$ between the $EW_{th}$ evaluated from the synthetic models with rotation ($V$ $\sin{i}$ = 250 Km/s) and the observed ones. In this analysis we excluded the He-weak stars. The following results are obtained:

\begin{figure}
	\includegraphics[width=\columnwidth]{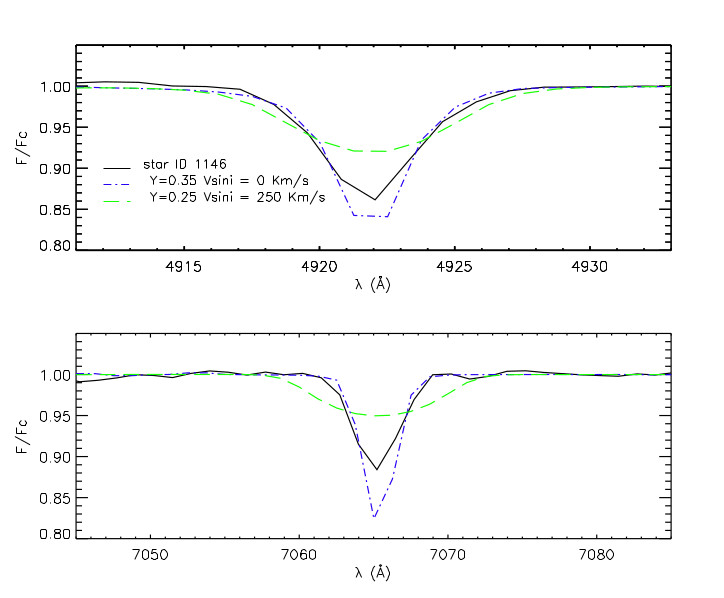}
\caption{ Spectra of the star \#1146 (black line), in the range of wavelength around the He lines at $\sim$ 4922 $\AA$ and 7065 $\AA$. It is compared with the synthetic spectra calculated with Y=0.35 and $V\sin{i}$=0 Km/s (blue dot-dashed line), and with Y=0.25 and  $V\sin{i}$ =250 Km/s (green dashed line). }

    \label{specrot}
\end{figure}
 
 \begin{enumerate}

 \item $\Delta_{EWrot}$ = 0.04 $\pm$ 0.04 for NGC\,1850 A 
 \item $\Delta_{EWrot}$ = 0.03 $\pm$ 0.05 for NGC\,1850 B 
 
 \end{enumerate}

 In a fully similar way and including the same stars, we derive the mean difference between the $EW_{th}$ estimated from the models with V$\sin{i}$=0 but Y different (Y=0.25 for the normal stars, Y=0.35 for the He-enhanced, but Y=0.27 for \#1069), and the observed ones.
In this case,  we found  the following values:
\begin{enumerate}

\item $\Delta_{EW}$ = 0.0007 $\pm$ 0.03 for NGC\,1850 A 
\item $\Delta_{EW}$ = 0.002 $\pm$ 0.050 for NGC\,1850 B

\end{enumerate}

The comparison of the values of this indicator ($\Delta_{EW}$ $<<$ $\Delta_{EWrot}$) shows that the synthetic models without rotation provide a fit of the observed EWs of much better quality than the synthetic models with rotation.
This can also  be appreciated by eyes in Fig. \ref{specrot}, where we show the spectra of the NGC\,1850 B star \#1146 (black line) in the range of wavelength around two He I lines  at 4922 $\AA$ (upper panel) and 7065 $\AA$ (lower panel). The two synthetic spectra are also plotted, they represent the results of the models with rotation and Y=0.25 (green dashed line), and  without rotation but Y=0.35 (blue dot-dashed line) . From this plot, it is evident that the model with rotation does not reproduce the wings and the depth of the He lines.
Instead the  He-enriched models fit well the wings of the lines, but are too deep. 
It is important to recall here that the resolution of the MUSE spectra is quite low and prevents a strong  conclusion.
A further intriguing possibility that could be confirmed or rejected 
by the analysis of high resolution spectra is briefly reported. Since fast rotation induces turbulent diffusion in the stellar interior,  which drives the CNO-cycled material from the core to the envelope \citep{meynet00}, a possible hypothesis could be  an He-enhanced (less than 0.35 in mass fraction) due to a stellar rotation (less than 250 Km/s).

\section{Conclusions}

In this paper we have examined for the first time the He abundance of B stars in the young ($\sim$ 90 Myr) binary LMC cluster NGC\,1850. 
 This system could be a unique bridge between two young massive stellar clusters observed during their process of formation of a multiple population and the old GCs exhibiting multiple populations. 

We analysed the spectra of 20 B stars extracted by MUSE cubes.

To determine the He abundance, we compared the EWs of four He lines (4922 \AA\, 5015 \AA, 6678 \AA\ and 7065 \AA) calculated from the MUSE spectra with the ones found analyzing the theoretical spectra.

These spectra has been computed with the code $SYNSPEC$, using non-LTE line-blancked model atmospheres of the grid $BSTARS2006$ for early B-stars \citep{Lanz07}.
We computed the model considering four different  He mass fraction : 0.25, 0.27, 0.30, 0.35.

The results can be summarized as follows:
\begin{itemize}
\item We found a not homogeneous He abundance. In particular, we can divide the targets of our sample in three group: He-normal  (Y=0.25 $\pm$ 0.01),  He-weak  (Y < 0.24) and He-rich (Y> 0.26);
\item The last group is intriguing because all stars, but one,  belong to the young isochrone, at 15 Myr,  4 of them are at a distance less then 5" from the center of NGC 1850 B, the other are in the edge between the two clusters.
 The mean value of the He mass fraction is about of 0.35 $\pm$ 0.02.

 \item Some He-rich stars are in common with the weak-O stars found by \cite{antony22}. If this is confirmed, e.g through the analysis of high resolution spectroscopy, it will be the first prove of  anticorrelation in  massive cluster in the MCs younger than 2 Gyr.

 \item The He -normal and  -weak  stars are associated to the isochrone at 90 Myr.

 
\item We evaluated the effect of the rotation on the He abundance by computing synthetic spectra with Y=0.25 and V$sin$ $i$  = 250 Km/s, in agreement with the works of \cite{kamann23} and with  the $v_{eq}$ values of the best  isochrone that fit the CMD. 




 


From our qualitative analysis,  the models without rotation but different Y fit better the EWs observed of NGC1850 B stars, but the resolution of the MUSE spectra is too low to have a strong  conclusion. 

\item We highlight that unfortunately our sample of He-normal and He-rich stars do not 
overlap in effective temperature, thus hampering to quantify 
possible source of systematic errors. 

\end{itemize}

In order to gain a deeper understanding of the characteristics or properties of the He features and clarify their nature, it is mandatory to perform high-resolution spectroscopic analyses (on which we will focus in future observational campaigns).

\section*{Acknowledgements}
RC thanks Antonio Sollima,  dearest friend, colleague and creator of this work.
Sadly he passed away prematurely, before to see the final results.  She thanks him for his advises and supports, but even more for his friendship, his joy, his humour, his music, his chats. She will be forever grateful to have had the opportunity to work together and to spent part of her life  in his company.\\
This work is based on observations collected at the European Organisation for Astronomical Research in the Southern Hemisphere under ESO program 0102.D-0268(A) and  made with the NASA/ESA
Hubble Space Telescope and obtained from the Hubble Legacy
Archive, which is a collaboration between the Space Telescope
Science Institute (STScI/NASA), the Space Telescope European
Coordinating Facility (ST-ECF/ESA) and the Canadian Astronomy
Data Centre (CADC/NRC/CSA).\\
We thank the anonymous referee for the valuable comments and suggestions that improved the quality of the publication.

\section*{Data Availability}
The data underlying this article will be shared on reasonable request to the corresponding author.



\bibliographystyle{mnras}
\bibliography{biblio} 





\bsp	
\label{lastpage}
\end{document}